# Lipid Controlled Non-Monotonic Assembly and Rheology of an Egg Yolk Protein at Water-Soybean Oil Interface

Nancy Jaglan[1], Rumal Singh[1], and Sajal K Ghosh[1*]

[1]*Department of Physics, School of Natural Sciences, Shiv Nadar Institution of Eminence, NH-91, Tehsil Dadri, Gautam Buddha Nagar, Uttar Pradesh-201314, India*

## Abstract

An essential component of food items like mayonnaise and salad dressing is hen egg yolk. Phosvitin (PVT) is a phosphoprotein, which exists in the granules of this hen egg yolk which stabilizes the food emulsion by preventing phase separation. To understand, how this protein is adsorbed at the interface of water and edible oil in the presence and absence of lipids is essential for improved control in food production. To monitor the kinetics of this adsorption, the dynamic interfacial tension has been determined in the current investigation. The drop in interfacial tension over time indicates the adsorption of protein at interface which is enhanced on increasing the concentration of protein in water phase. However, at higher concentration, the positive activation energy hinders the adsorption process resulting a saturated interfacial tension. The dilation rheology of the macromolecular film at this oil-water interface shows the elastic nature of the film to be greater than the viscous nature, indicating the formation of a soft gel film. At low concentration, the zwitterionic lipid, 1-palmitoyl-2-oleoyl-sn-glycero 3-phosphocholine (POPC), promotes this protein adsorption at the interface. Interestingly, at high concentration, the lipid overtakes the interface removing the protein from there. The lipid-protein composite film again shows the nature of a soft gel. The non-monotonic effects of lipids on assembly of protein at the water-edible oil interface is an important observation to optimize the composition of relevant food products.

*Email of corresponding author: sajal.ghosh@snu.edu.in

## 1. Introduction

Two or more immiscible liquids, usually an aqueous and oil phase, are used to create emulsions. Food emulsions, which fall under the category of macroemulsions, are typically unstable as oil droplets start to get closer to one another to produce, eventually, larger droplets. Such a large droplet separates from the aqueous phase when it is big enough resulting in complete phase separation [1]. A number of physical and chemical parameters, including the type of fluids, the droplet size and the mixing technique [2], can affect how quickly this process proceeds. This phase separation kinetics can be slowed down or hindered by using emulsifying agent.

The common oil-in-water emulsion known as mayonnaise is made from hen egg yolk, oil, water, vinegar, salt, and lemon juice. [3], [4]. Egg yolk acts as a natural emulsifying agent due to the presence of biomolecules such as phospholipids and proteins that can act as surfactants [2, 5]. These molecules may interact with the aqueous and oil phases since they have both hydrophilic and hydrophobic moieties. They prevent droplet coalescence and inhibit phase separation by adsorbing at the oil–water interface, thereby lowering interfacial tension and creating a barrier around dispersed droplets [6]. Therefore, to formulate a stable product, it is important to quantify the kinetics of adsorption of molecules to the oil-water interface and investigate the influence of one on assembly of other. Further, the assembled layer at the interface has an important role in deciding the overall rheological behaviour of the product. Therefore, characterizing this behaviour of the interface is essential. This work presents both these aspects by investigating the interfacial activity of egg protein phosvitin (PVT) at oil-water interface both with and without of an egg yolk phospholipid.

PVT is highly phosphorylated phosphoprotein , which together with high-density lipoproteins (HDL), constitutes the granular fraction of hen egg yolk. It contains many negatively charged amino acid groups as it is strongly phosphorylated. This protein is an effective emulsion stabilizer due to its amphipolar nature. It accounts for approximately 25% of yolk granule proteins and is characterized by an exceptionally high serine content, the majority of which are phosphorylated [7]. Several PVT polypeptides have been sequenced, revealing that the protein can be structurally described as extended core rich in negatively charged phosphoserine residues, along with an approximately 15-amino acid C-terminal region that is comparatively rich in hydrophobic residues [8,9]. Egg yolk contains major phospholipid POPC [10], [11], which are described by presence of both hydrophilic and hydrophobic regions which enable them to assemble at the oil-water interface to enhance the emulsions stability.

There are a few reports investigating the combined lipid and protein system at the interface, showing how lipid composition influences protein conformation, aggregation, and organization [12]. Havela et al. demonstrates that adsorbed proteins can reorganize surrounding phospholipids, creating lipid-rich domains that influence further protein adsorption and assembly [13]. The work presented how phospholipids can either promote or inhibit protein adsorption depending on concentration, adsorption kinetics, and interfacial packing. Others explain how proteins unfold, reorganize, and self-associate after encountering lipid-containing interfaces [14]. More detailed and systematic studies are required in this research field to figure out a general mechanistic framework for understanding protein assembly driven by associated lipids which are basic components of numerous food products.

In the present study, the assembly of the protein, lipid and their mixture at oil-water interface was quantified by the dynamic interfacial tension with pendant drop tensiometer. Further, the kinetics of their adsorption was investigated as a function of concentration of the components. It was striking to observe that, at a low lipid concentration, the lipids help the protein to assemble at the interface, then they overtake the interface by assemble themselves at higher concentration. The dilatational rheology method was executed to follow the viscoelastic nature of interface after formation of a stable macromolecular layer. The interface shows a soft gel nature where value of storage modulus is higher compared to the loss modulus.

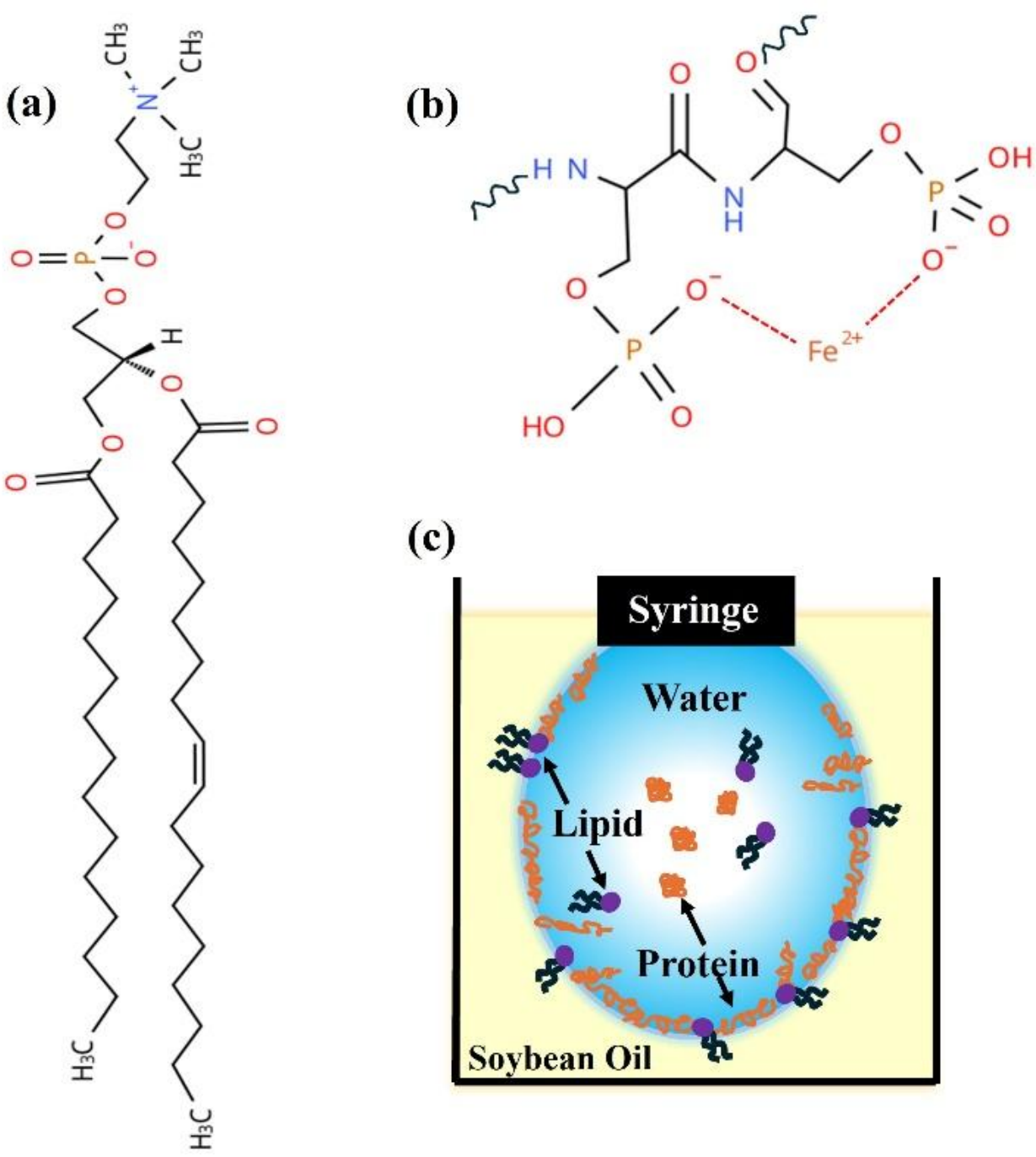

*Figure 1: Chemical structure of (a) zwitterionic lipid POPC and (b) egg yolk protein phosvitin (PVT). (c) Schematic of a lipid and protein dispersed water droplet immersed in a medium of soybean oil.*

## 2. Materials and Methods:

### 2.1 Materials:

The zwitterionic lipid, 1-palmitoyl-2-oleoyl-sn-glycero 3-phosphocholine (POPC) (Fig.1(a)) was purchased in powder form from Avanti Polar Lipids (Alabaster, AL). Phosvitin (PVT) (Fig.1(b)), an anionic phosphoprotein in hen egg yolk, with molecular weight of 35 kg/mol, and soyabean oil in light yellow colour were purchased from Sigma-Aldrich. The solutions of PVT and POPC were prepared in deionised water (Milli-Q).

### 2.2 Dynamic Interfacial Tension

The interfacial tension has been measured with an optical tensiometer (Biolin Scientific, Theta-Flex) . In this technique, a pendant droplet of volume 7 µL of aqueous solution of lipid, protein or their mixture was immersed in medium of soybean oil taken in a quartz cuvette (Fig. 1 (c)). The shape of the drop was monitored with a high-resolution camera capturing the image with 1 to 20 fps. This technique is widely recognised as a reliable approach for studying the time-dependent behaviour of interfacial tension in systems involving surfactants, proteins, and nanoparticles.[15,16, 17]. The Young-Laplace equation was fitted to the resulting pendent drop profile to find the interfacial tension [18]. After formation of pendant drop, the interfacial tension was monitored for 12 hr at a constant temperature of 25°C.

### 2.3 Dilatational Rheology

After the interfacial tension reached an equilibrium, dilatational rheology measurements were performed at the oil–water interface with the lipid, protein, or their combination. The interfacial area, $a(t)$, of the pendent drop was varied sinusoidally and, then the corresponding interfacial tension, γ(t), was recorded. Here, an area amplitude of variation was taken to be 0.5%, which is calculated from $\frac{\Delta A}{A_0}\times 100\%$, where $A_o$ is the initial interfacial area of the droplet and $\Delta A$ is the variation in $A_o$. The viscoelastic property of the nano layer formed at oil-water interface was determined from the phase lag ($\varphi$) between applied strain, droplet area and measured stress, the interfacial tension. The parameters $G'' = \Delta\gamma \frac{A_0}{\Delta A} cos\,\varphi$ and $G'' = \Delta\gamma \frac{A_0}{\Delta A} \sin\phi$ represent dilation storage and loss moduli, respectively [19, 20, 21, 22]. Here, $\Delta\gamma$ is the amplitude of oscillation of interfacial tension. The oscillation was varied from 50 to 300 mHz with an increment of 25 mHz.

## 3. Results and Discussion

To identify protein assembly at soybean oil-water interface, the dynamic interfacial tension was studied. Once the interface reached to an equilibrium with complete coverage by protein, lipid or their mixture, the interfacial rheology was performed to characterise the viscoelastic nature of the molecular film. For a fixed concentration of the protein PVT, how the different concentrations of lipid, POPC affects its assembly was also studied.

### 3.1 Assembly of Phosvitin (PVT) Protein at the Interface

***3.1.1 Dynamic Interfacial Tension :*** The interfacial tension ($\gamma$) of oil-water interface with PVT dispersed in water droplet at varying concentration was measured over a period of 12 hrs (Fig. 2(a)). For all concentrations, PVT molecules adsorb at oil–water interface, leading to a progressive decrease in magnitude of $\gamma$. An increasing concentration of protein leads to greater assembly at the interface, thereby causing a more pronounced reduction of $\gamma$. Figure 2(b) presents these values at different PVT concentrations measured after 10 hrs of immersing the water droplet in the oil medium. The increasing PVT concentration from 0.1 to 0.5 g/L, drops $\gamma$ from $17.89 \pm 0.07$ to $11.63 \pm 0.18$ mN/m. The concentration beyond 0.5 g/L does not cause any further change in $\gamma$ indicating a full coverage of the interface by the protein molecules. Even if there are a few molecules assembled beneath the interfacial layer because of protein-protein interaction, they may not contribute to the value of $\gamma$.

The diffusion mechanism of PVT was analysed using the modified Ward–Tordai model [23]. This equation defines the time-dependent adsorption density of diffusing molecules at the interface, including the diffusion as well as the back-diffusion process. The adsorption density ($\Gamma$) of molecules at the interface with diffusion coefficient ($D$) at any time ($t$), is governed by

$$\Gamma\,(t) = \left(2\sqrt{\left(\frac{D}{\pi}\right)}\;\left(c_o\,\sqrt{t}\;-\;\int_0^{\sqrt{t}} c\;(0, t-\tau)\;d\,\sqrt{t}\right)\right) \qquad (1)$$

where $c_o$ represents bulk concentration, and c is the time-dependent concentration at interface. Here, first term relates to the diffusion mechanism, while the second term is associated with the back-diffusion. Unfortunately, the back-diffusion term is unknown explicitly due to an unknown form of integrant. For the short-time approximation, $t \to 0$, the equation can be simplified with the help of the Gibbs adsorption equation in terms of $\gamma(t)$ as follows [24],

$$\gamma(t) \;\cong \gamma_0 - 2RTc_o\sqrt{\frac{Dt}{\pi}}\;, \qquad (t \to 0) \qquad (2)$$

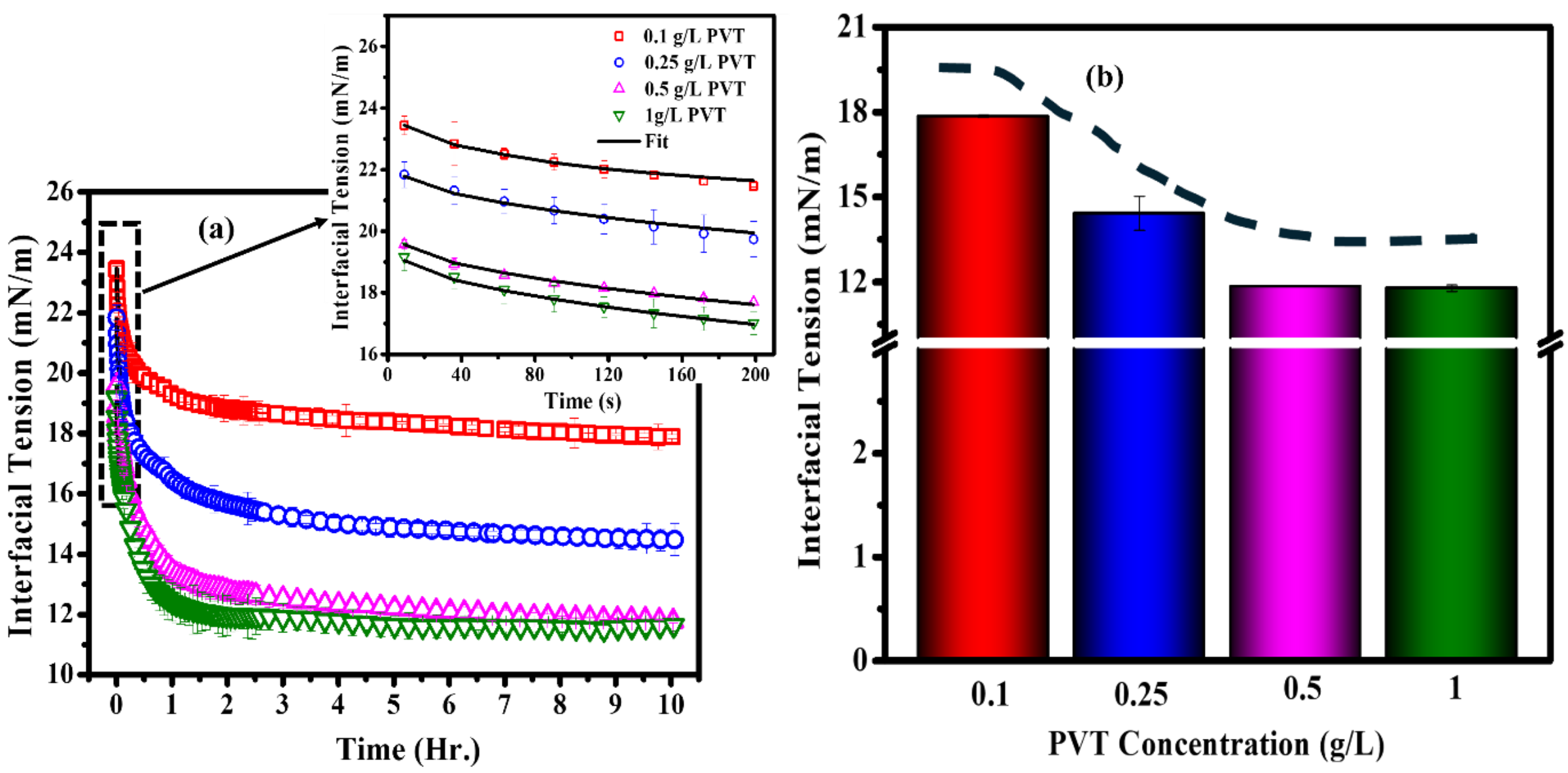

*Figure 2 (a) Dynamic interfacial tension (γ (t)) of soybean oil-water interface on the assembly of phosvitin (PVT) protein. The protein concentration was varied from 0.1 to 1 g/L. (b) Saturated value of the tension after 10 hrs of immersion of the PVT dispersed water droplet into the oil. The inset of (a) shows fitting of the data with asymptotic relation of Ward -Tordai equation (t→0 region).*

Here, $\gamma_0$ is the interfacial tension for pristine interface, T is the temperature, and R is the universal gas constant. At this approximation, protein adsorption kinetics are expected to be governed by diffusion-controlled processes. The diffusion coefficient ($D$) of PVT has been determined by fitting the dynamic interfacial tension data (inset of Fig. 2(a)) with the values found to decrease with increasing protein concentration ($c_o$) in the water droplet as shown in Fig. 3(a). The diffusion coefficient of the protein in bulk water was also calculated to be $D_{\text{Stokes}} = 5 \times 10^{-11}$ m$^2$/s, (black dashed line in Fig.3(a)) using the Stokes–Einstein equation [25].

$$D_{\text{stoke}} = \frac{K_B T}{6\pi\eta R_g} \tag{3}$$

where $K_B$ is the Boltzmann constant, $\eta$ is viscosity of aqueous solution with $R_g$ being the radius of gyration of PVT, which is estimated to be ~5 nm. From this calculated value and the measured values of diffusion coefficients, it is evident that the increasing concentration shifts the adsorption from diffusion-controlled to interaction-controlled, where the protein molecules begin interacting with each other near the interface. Thus, the motion of proteins around the interface involves two mechanistic steps: the diffusion-controlled motion from t bulk solution to the sub-layer at interface and overcoming the potential barrier (ΔE) between the real interface and the bulk phase. Consequently, as explained by Miller et. al, the kinetics for adsorption of the protein at interface may have three regimes: the diffusion-controlled initial regime, the energy barrier-controlled final regime and the mixed diffusion–barrier-controlled intermediate regime [26]. A relation of effective diffusion coefficient ($D_{t\to 0}$) with the activation energy or potential barrier (ΔE), including the diffusion coefficient ($D_{stoke}$) obtained from the Stoke-Einstein equation (3), can be given by the following expression[40],

$$D_{t\to 0} = D_{stoke} \exp\left[-\frac{\Delta E}{K_B T}\right] \tag{4}$$

Here, this equation is utilized to quantify the activation energy, which is displayed in Fig. 3(b) as a function of PVT concentration. This is the energy expected for protein to adsorb at interface with corresponding rearrangement after adsorption. The adsorption kinetics of the protein is diffusion-controlled with a negative activation energy up to the concentration of 0.25g/L, above which the energy becomes positive. This means that at lower concentrations, the molecules are spontaneously assembled at the interface. However, at higher concentrations, they are repelled by the molecules that are already assembled at interface, causing the hindrance to further molecules to absorb.

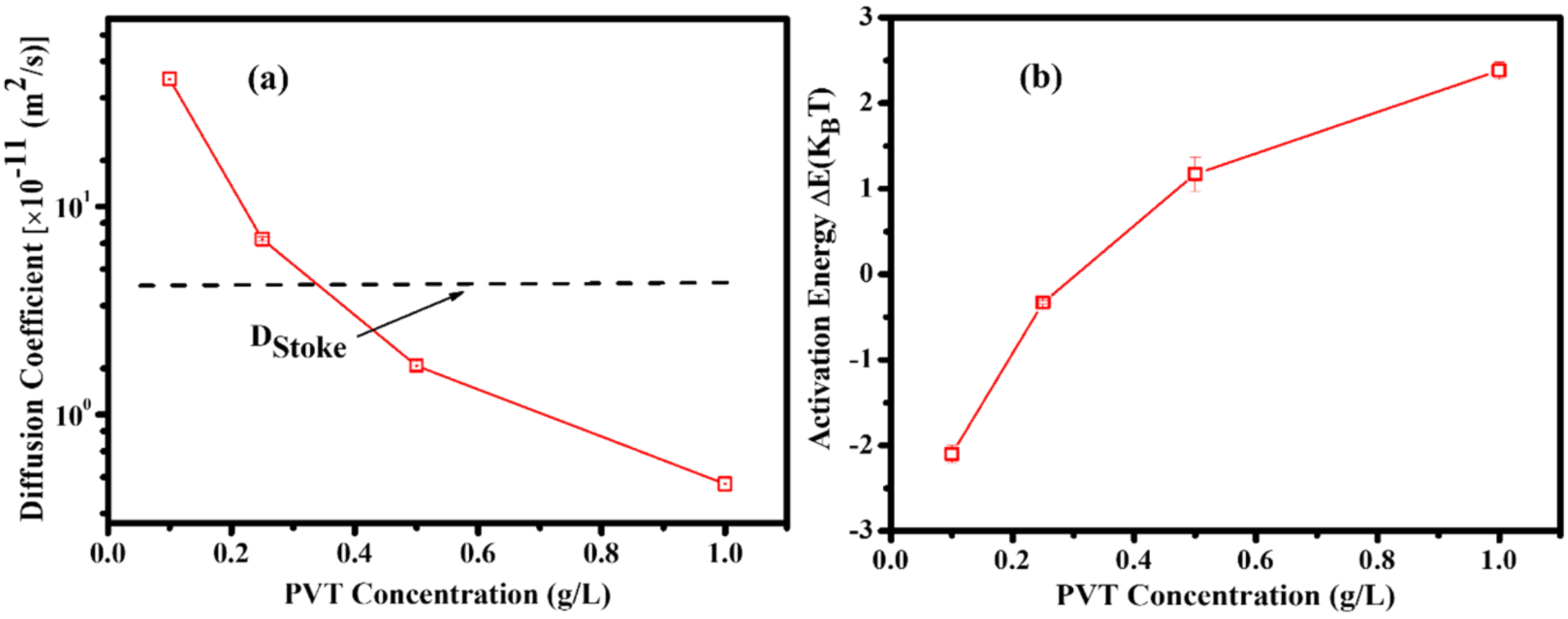

*Figure 3: (a) Variation in diffusion coefficient (D) of phosvitin (PVT) protein determined by fitting the data in Fig. 2 (a) in the region $t \to 0$. The black dashed line shows the value of D calculated from Stoke-Einstein equation. (b) Activation energy for different concentration of the protein.*

Gravidel et al. have investigated the activation energy for different concentrations of a polymer, Sulfethylated Kraft Lignin (SEKL) at the liquid-liquid interface [27]. At the water-cyclohexane interface, they studied that adsorption kinetics of SEKL are only diffusion controlled for the concentration of 0.25-0.5 wt%. At higher concentrations, an energy barrier exists, resulting in the magnitude of activation energy in the range of 4.5 to 5.7 $K_BT$. Kutuzov et al. have calculated the energy barrier of about a few $K_BT$ for nanoparticles [28]. The energy found for the present system is quite similar to these reported systems. For mayonnaise, the emulsion stabilised by egg protein and other surface-active molecules like lipids, this activation energy may help to explain and predict the emulsion stability, texture, and processing behaviour [4].

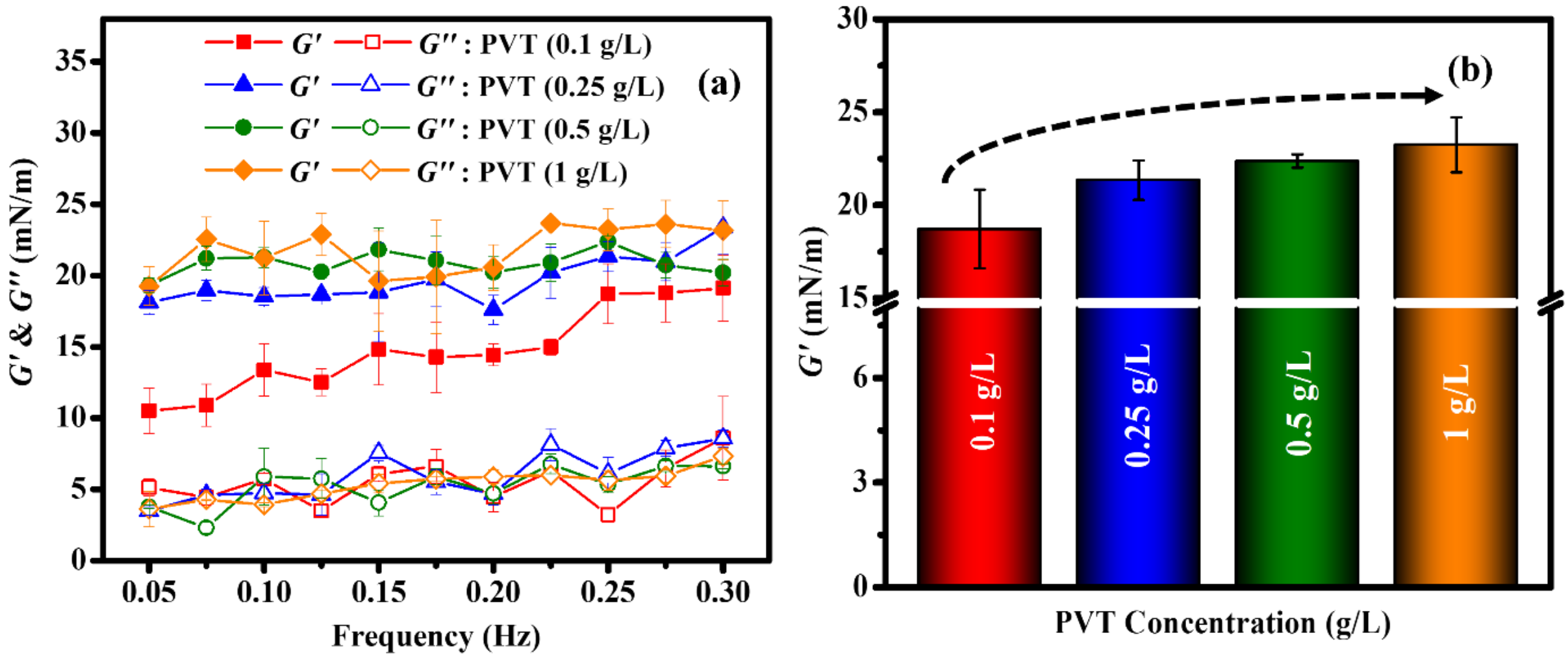

*Figure 4: (a) Variation in viscoelastic behaviour at different concentrations of phosvitin (PVT) accumulated at interface of water and soybean oil plotted as function of frequency. (b) Values of $G'$ for the assembled film at a particular frequency of 0.25 Hz.*

***3.1.2 Viscoelastic Nature of Assembled PVT Film at Interface:*** The viscoelastic behaviour of PVT molecules assembled at oil-water interface is shown in Fig. 4(a) with the values of storage ($G'$) and loss ($G''$) moduli plotted with frequency. It is noted that these measurements were performed once the molecules saturate the interface with a final value of the interfacial tension (Fig. 2(a)). The magnitude of $G'$ of the film is higher than $G''$ for all the frequencies of oscillation, suggesting the film to be elastic in nature, which is predominant over its fluidic nature. However, the low value of $G'$ suggest the assembled film to be a disordered solid exhibiting a two dimensional (2D) soft gel state [29, 30]. Such observation has been reported for other biopolymeric systems, such as Xanthan-gum [31] and Pea Protein Isolate (PPI) [32]. The interfacial film strength and elasticity of PVT are within a similar range when these results are compared to those obtained for PPI. With an increase in the concentration of the PVT, there are more molecules to assemble at the interface, which makes the layer to require higher energy to deform exhibiting an enhanced value of $G'$. At *0.25 Hz,* the value increases from 18.73 to 23.25 mN/m with increase in concentration from 0.1 to 1 g/L (Fig. 4(b)). It is worth of noticing that even the magnitude of γ does not drop beyond 0.5 g/L, the value of $G'$ still increases, though slightly, at higher concentration. The secondary layer which assembled at higher concentration beneath the primary layer at the interface may not contribute to drop the interfacial tension but the deformation of primary layer at the interface will be influenced by the secondary layer because of protein-protein interaction of the layers. This rheological behaviour of PVT is particularly interesting to stabilize the emulsion and modify the texture, in food products .

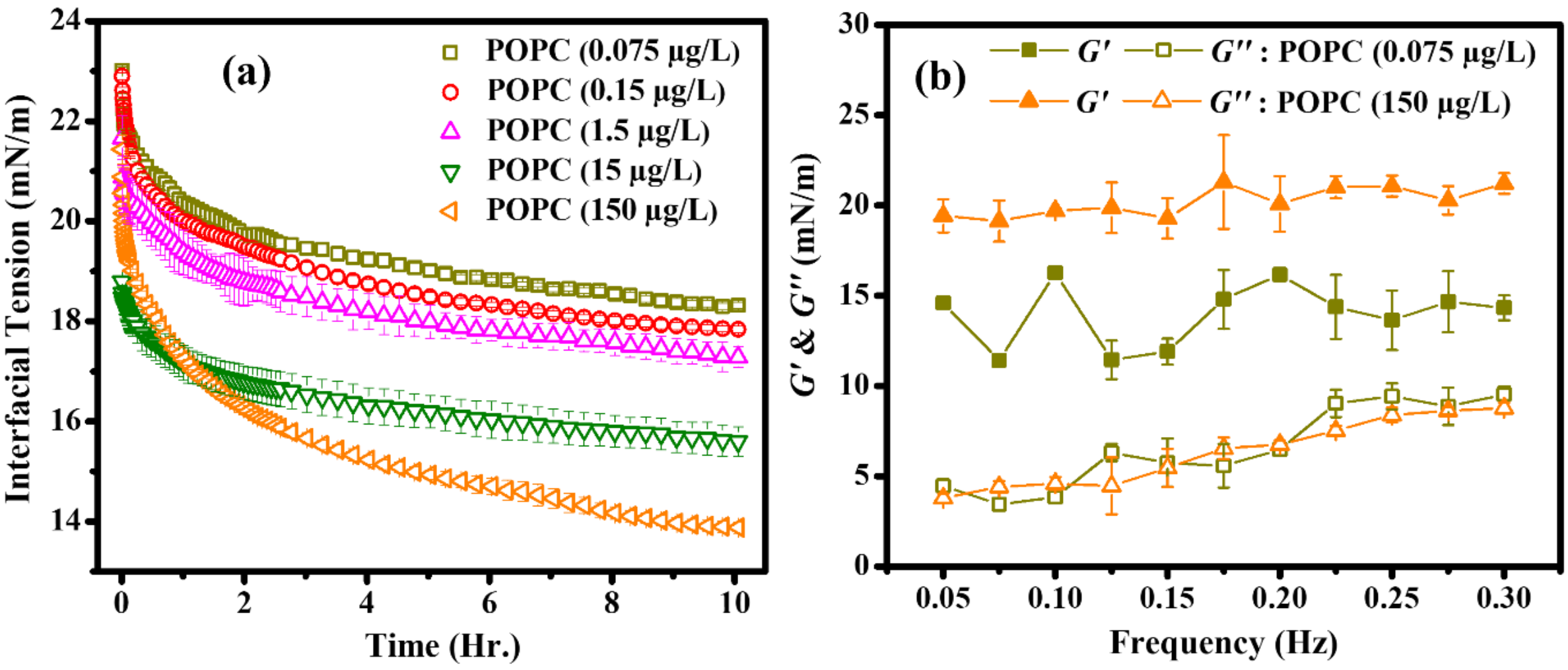

*Figure 5: Figure of (a) dynamic interfacial tension at different concentrations of POPC for 10 hrs.(b) Variation of viscoelastic behaviour of POPC concentration at the oil-water interface plotted as function of frequency.*

### 3.2 Assembly of Zwitterionic Lipid POPC at Oil-water Interface

Even though the adsorption of zwitterionic lipids at oil-water interface is well studied [33], to understand the effect of such a lipid on the assembly of PVT protein, the nature of POPC assembly is revisited. Here, the varied concentration of lipid was taken into water droplet which was then immersed into soybean oil. Their assembly at the interface considerably drop the values of $\gamma(t)$ which is shown in Fig. 5(a), exhibiting a saturated value of 18.32 mN/m at the lowest lipid concentration of 0.075 μg/L. As observed for protein, the lipid also exhibits a higher drop in the value of $\gamma$ at higher concentration, as shown in Figure 5(a). Walker et al. have compared the interfacial tension for fixed concentrations of different PC lipids [34]. DSPC shows a lower drop in surface tension from 44.5 to 42.6 mN/m, then for DPPC, DMPC and DLPC, the final interfacial tension is 40, 35 and 31 mN/m, respectively.[34.]. Compared to all these lipids, POPC indicates a higher surface activity as it drops the interfacial tension to a much lower value. Following the analysis discussed in the previous section, the diffusion coefficient, $D_{t\to 0}$, of the lipid gives $6 \times 10^{-8}$ $m^2$/s at the concentration of 150μg/L which is comparable to the values reported earlier [33]. The viscoelastic behaviour of lipid film at oil-water interface with frequency is displayed in 5(b). The higher value of $G'$ represents the elastic behaviour of the lipid film over the fluidic behaviour. As observed for protein, with increasing concentration, the film requires higher energy to deform, resulting in an enhanced value of $G'$.

### 3.3 Lipid Controlled Assembly of Phosvitin (PVT) Protein at Interface

***3.3.1 Dynamic Interfacial Tension:*** To investigate the impact of this POPC phospholipid on assembly of the PVT protein at interface, the protein and lipid were mixed in the water droplet which was then immersed into the oil. To measure the value of $\gamma$, the protein concentration (P) was fixed to be 0.25 g/L while the concentration of the lipid (L) was varied from 0.075 to 150 μg/L. It is interesting to observe that the assembly of the protein does not show a monotonic behaviour with the concentration of the added lipid in the mixture (Fig. 6). Low lipid concentration promotes the adsorption of protein at interface, resulting in greater drop in the value of $\gamma$ compared to the pure protein. For example, the individual saturated values of $\gamma$ after 10 hrs are 14.5 mN/m for 0.25 g/L for protein (Fig. 2(b)) and 18.32 mN/m for 0.075 μg/L for lipid (Fig. 5 (b)), which drops to a saturated value of $\gamma$ to 12.8 mN/m (Fig. 6 (b)) when they are mixed. At neutral pH, which is the case in the present study, the protein behaves as an anionic polymer because of the presence of phosphoserine groups. In case of pure protein, after assembly of few molecules, the interface may become negatively charged, which may repel further protein to assembly at the interface. However, the positively charged choline group of the lipid may screen such a repulsion, promoting more protein at the interface. This hypothesis requires more systematic studies to comprehend the interface better. The promoting effect of the lipid diminishes for the concentration from $L_1$ to $L_3$, as displayed in Fig. 6(b). At higher concentrations of the lipid, the interface is predominantly occupied by the lipid as depicted for the concentrations of $L_4$ (15 μg/L) and $L_4$ (150 μg/L). Such a non-monotonic adsorption of protein with different lipid concentrations has been recently reported by Pontillo et al. by an MD and MetaD simulation work [35]. They investigated that phospholipids exert a concentration-dependent influence on protein adsorption and conformation. At low concentrations, they promote protein adsorption by inducing partial unfolding and increasing the radius of gyration, thereby exposing hydrophobic regions (Fig. 8). In contrast, at higher concentrations, phospholipids preferentially occupy the interface and form a steric barrier that limits protein adsorption. Moreover, while phospholipids dispersed in solution favour expanded protein conformations, phospholipid-rich interfaces induce more compact protein structures through interfacial confinement and specific lipid–protein interactions. The diffusion coefficient (D) and the activation energy have not been extracted for the mixt system as the implemented equation is suitable only for single-component system. A careful attempt is required to find these parameters for such a system which will be a subject of future work.

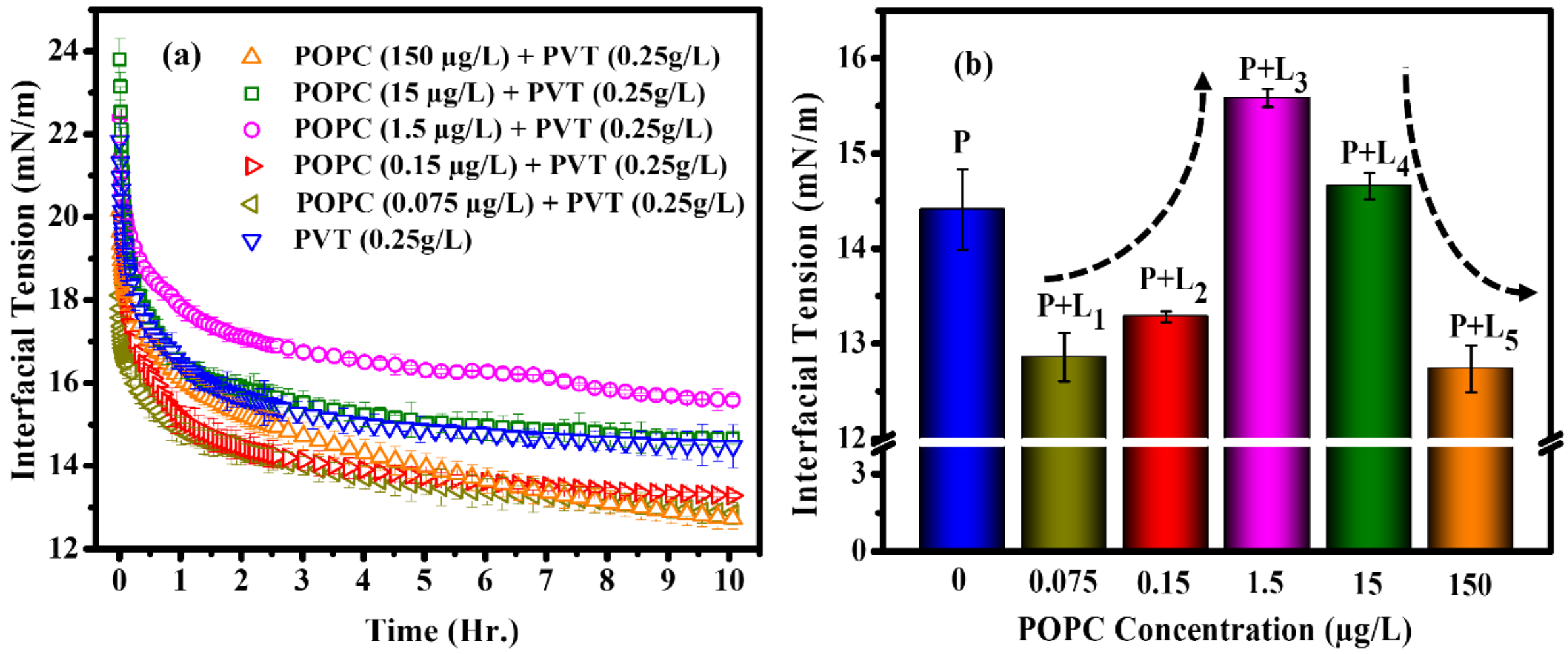

*Figure 6: (a) Dynamic interfacial tension of soybean oil-water interface during the assembly of phosvitin protein in the absence and presence of POPC lipid. (b) Saturated value of the tension after 10 hrs of immersion of the PVT (P) or mixture of PVT and lipid (P+L) dispersed water droplet into the oil. The varied concentrations of lipids are denoted as $L_1$ (0.075 µg/L), . . . , $L_5$ (150 µg/L) with a fixed concentration of the protein (P) of 0.25 g/L.*

***3.3.2 Viscoelastic Nature of Protein-lipid Film at Interface:*** The viscoelastic behaviour of PVT molecules with and without POPC molecules assembled at oil-water interface with frequency range from 0.05 to 0.3 Hz is shown in Fig. 7(a). The interfacial layer is more elastic in nature, which is shown by a greater value of $G'$ as compared to $G''$ for all the solutions. For a fixed concentration of PVT (P), the POPC concentration $L_1$ and $L_2$ show a lower value of $G'$ which is probably due to the unfolding nature of the PVT molecules in the presence of POPC as predicted by the simulation work discussed above [35]. Suh a layer with unfolded protein molecules makes the layer less rigid. For the concentration $L_3$, POPC molecules start to accumulate at the interface along with the PVT molecules. The presence of POPC molecules makes the layer more elastic. Further increase in the POPC concentration ($L_4$ and $L_5$) causes more POPC molecules to accumulate at the interface, while the PVT molecules go into the bulk causing further increase in value of storage moduli of the interfacial layer.

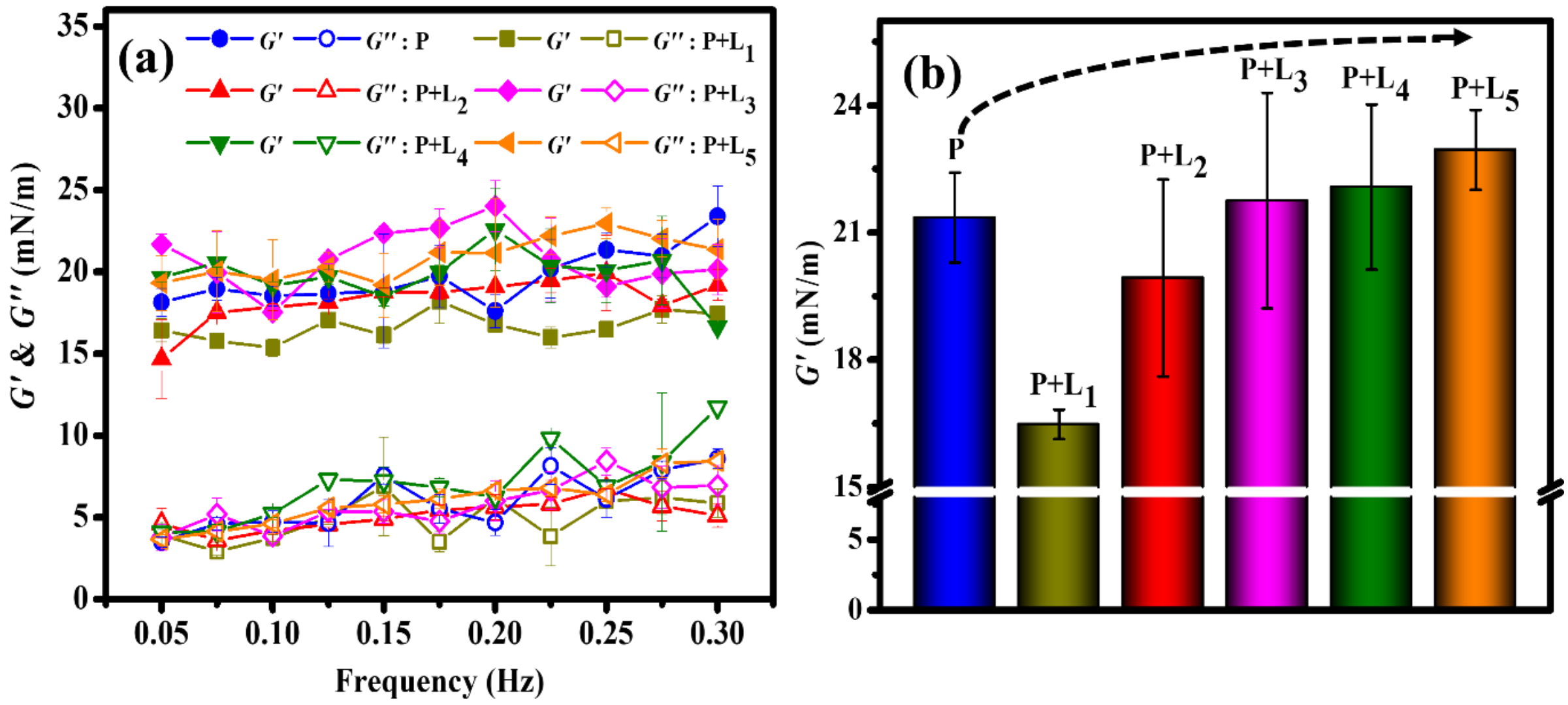

*Figure 7: (a) Variation of viscoelastic parameters with frequency for fixed concentration of PVT with and without different POPC concentration. (b) Bar graph represents G′ at frequency of 0.25 Hz. Here, P represents the PVT protein for 0.25 g/L concentration and L for the POPC lipid. $L_1$, $L_2$, $L_3$, $L_4$, and $L_5$ present 0.075, 0.15, 1.5, 15, and 150 µg/L of POPC, respectively.*

This study uses a simplified soybean oil–water interface, which may not fully represent the complexity of real food emulsions such as mayonnaise. The egg yolk contains a complex mixture of phospholipids dominated by phosphatidylcholine, along with phosphatidylethanolamine, sphingomyelin, phosphatidylinositol, and lysophosphatidylcholine. Therefore, in future, such a complex system needs to be considered to comprehend the molecular assembly at the interface. Further, present study only focuses on phosvitin protein, which can be replaced by other egg yolk proteins such as high-density lipoproteins (HDL), low-density lipoproteins (LDL), and apovitellenins, for a better understanding of the real food products. One interesting aspect here is the pH which can be examined as phosvitin is highly sensitive and may cause change in stability. Even though it is explained following the reported simulation work that at low concentration, the lipid helps unfolding the protein while at high concentration, they replace the protein from interface, no direct structural evidence is presented herein. This structural modification of protein and lipids at the interface can be explored using the high flux synchrotron-based X-ray scattering techniques. Such approaches will contribute to a mechanistic understanding of protein–phospholipid co-assembly and support the rational design of stable and functional emulsion-based systems.

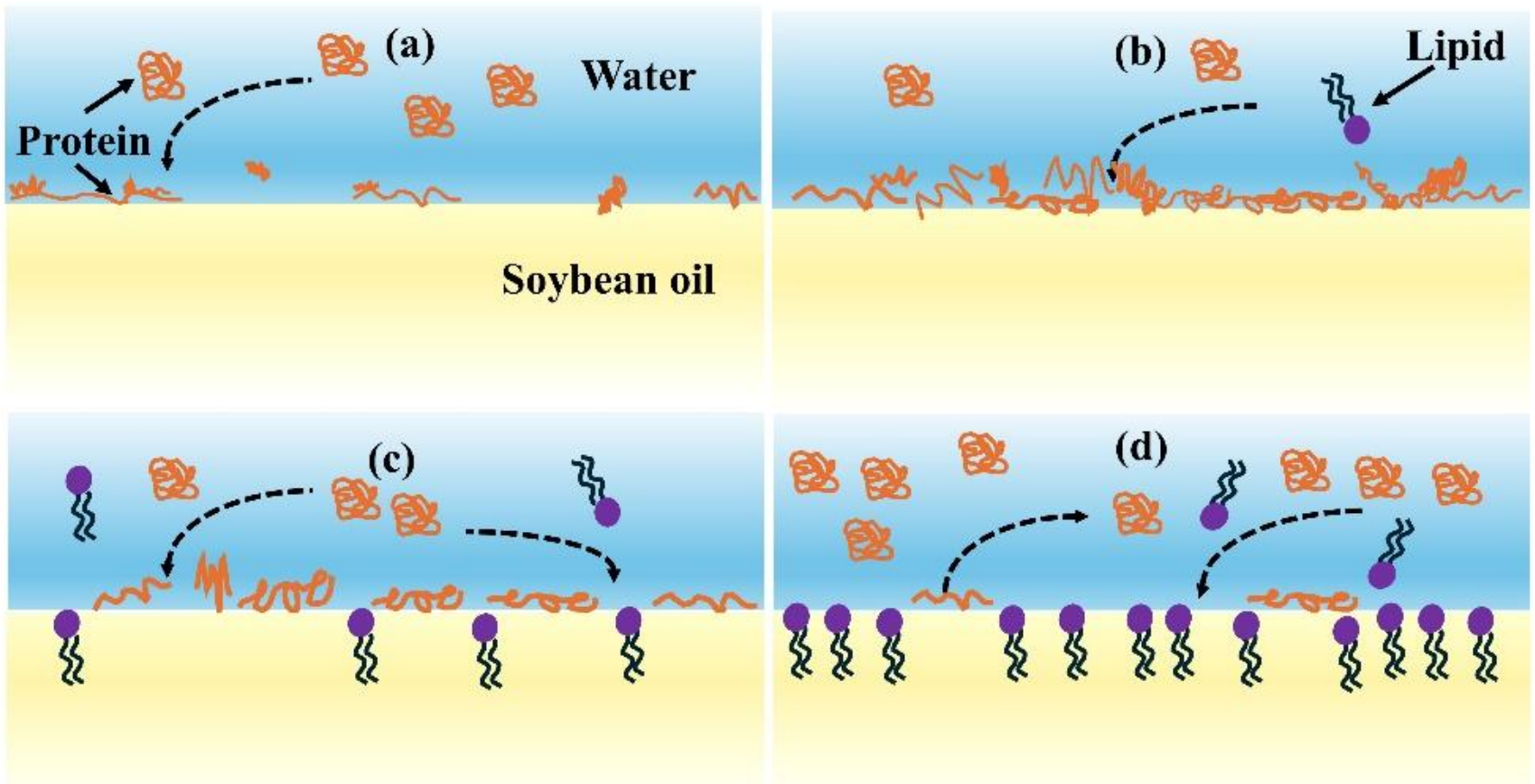

*Figure 8: Schematic of adsorption of protein at the soybean oil-water interface: (a) without lipid and (b, c, d) in the varying concentration of lipid for a fix concentration of protein.*

## 4. Conclusions

In this present work, the adsorption behaviour of an egg yolk protein and a phospholipid to soybean oil-water interface has been investigated. To comprehend the adsorption behaviour, dynamic interfacial tension was quantified for different concentrations of the protein phosvitin (PVT). The assembly of protein molecules at interface from the bulk changes its conformation as depicted in Fig. 8. The interfacial tension decreases with increasing PVT concentration, indicating the greater assembly of molecules at interface. The viscoelastic behaviour of interface is evaluated after interface is saturation with the molecular film. The results indicate that the viscoelastic nature increases with the protein concentration. Further, the decreasing diffusion coefficient with concentration indicates that the motion of molecules becomes slower. For the protein-lipid complex system, at low concentration, the lipid POPC promotes PVT adsorption by further reducing the interfacial tension and decreasing the storage modulus. With increasing concentration, POPC molecules start accumulate at interface and start to bring the PVT molecules back into the bulk.

## 5. Acknowledgement

The authors thank the Shiv Nadar Foundation for the financial support to conduct the research. The useful scientific discussion with Arshita Ishan is also acknowledged.